\documentclass[conference]{IEEEtran}
\ifCLASSINFOpdf
  \usepackage[pdftex]{graphicx}
  \DeclareGraphicsExtensions{.pdf}
\else
  \usepackage[dvips]{graphicx}
  \DeclareGraphicsExtensions{.eps}
\fi
\usepackage{amsmath}
\begin{document}
\title{Investigation on Radio Wave Propagation in Shallow
Seawater: Simulations and Measurements}
\author{
\IEEEauthorblockN{Eugenio~Jimenez, Gara~Quintana,\\
Pablo~Mena, Pablo~Dorta and \\Ivan~Perez-Alvarez}
\IEEEauthorblockA{
Institute for Technological Development\\ 
and Innovation in Communications (IDeTIC),\\ 
Universidad de Las Palmas de Gran Canaria, \\ 
Las Palmas, 35017, Spain. \\
e-mail: eugenio.jimenez@ulpgc.es}
\and
\IEEEauthorblockN{Santiago~Zazo, Marina~Perez}
\IEEEauthorblockA{
ETSI Telecomunicacion,\\ 
Universidad Politecnica de Madrid, \\
Madrid 28040, Spain.\\
e-mail: santiago@gaps.ssr.upm.es\\
e-mail: marina.perez@isom.upm.es}
\and
\IEEEauthorblockN{Eduardo~Quevedo}
\IEEEauthorblockA{
Oceanic Platform of the Canary\\
Islands (PLOCAN),\\ 
Telde 35200, Spain.\\
e-mail: eduardo.quevedo@plocan.eu}
}
\maketitle
\begin{abstract}
The authors present full wave simulations and experimental results of
propagation of electromagnetic waves in shallow seawaters. Transmitter and
receiver antennas are ten-turns loops placed on the seabed. Some propagation
frameworks are presented and simulated. Finally, simulation results are compared
with experimental ones.   
\end{abstract}
\IEEEpeerreviewmaketitle

\section{Introduction}
\IEEEPARstart{E}{lectromagnetic}
propagation through sea water is very different from propagation through air because
of water's high permittivity and electrical conductivity. Plane wave attenuation
is higher through water, and increases rapidly with frequency. With a relative 
permittivity of about $\epsilon_r$=80, water has the highest permittivity of any material and
this has a significant impact on the angle of refraction at the air/water 
interface. Conductivity of seawater is typically around 5S/m, while nominally 
fresh water conductivity is quite variable but typically in the mS/m range. 
Relative permeability is approximately $\mu_r$=1
so there is little direct effect on the magnetic field component but
conduction leads to strong attenuation of electromagnetic propagating waves.
\par
Another important consideration is the effect of the air-to-water interface. 
Propagation losses and the refraction angle are such that an electromagnetic 
signal can cross the air-to-water boundary and appears to radiate from an
antenna 
directly placed in the air above the transmitter. 
This effect aids communication from a submerged station to land and between 
shallow submerged stations without the need for surface repeater buoys. The air
path can be a key advantage. For example, if two divers are 1km apart at 2m 
below the surface, attenuation will be significantly less than anticipated from
the 1km through-water loss. 
\par
A similar effect is seen at the seabed, where its conductivity is lower than the
water one. The seabed is an alternative low-loss, low-noise,
communications path if both transmitter and receiver are placed on the seabed.
\par
\begin{figure}[!h]
\label{fig:geometry}
\centering
\includegraphics[width=8.5cm]{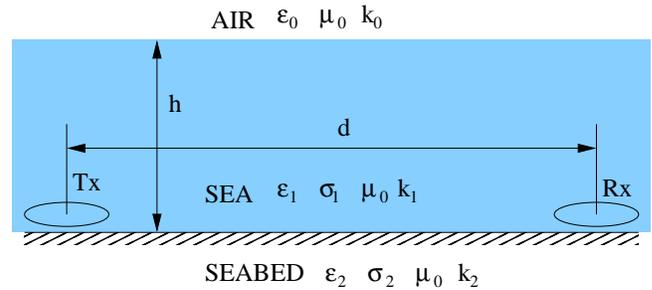}
\caption{Geometric configuration of testbed.}
\end{figure}

 
In many deployments a single propagation path will be dominant depending on the
placement of transmitter and receiver. In our case both transmitting and
receiving antennas will be placed on the seabed as it is illustrated in 
Fig~\ref{fig:geometry}.
\par
Full wave analysis of propagating EM waves in two-layers geometries was firstly
carried out by A. Sommerfeld at the begining of the XX century. Later, his work
was extended to multilayer geometries and 
in~\cite{king:1992}, \cite{wait:1996}~and~\cite{li:2009}
full wave analysis of geometries with two, three and more layers and 
applications are well summarized.

\section{Simulation frameworks}  
Along this communication, four different frameworks are going to be simulated.
\begin{IEEEitemize}
\item
Attenuation between two horizontal loops placed in free space.
\item
Attenuation between two horizontal loops inmersed in sea water.
\item
Attenuation between two horizontal loops placed on seabed to sea water
interface without air layer; two layers problem.
\item
Attenuation between two horizontal loops placed on seabed to sea water
interface with an air layer over sea water; three layers problem.
\end{IEEEitemize}
\par
In all cases, transmitter and receiver antennas are of the same kind: a 22cm.
radius ten turns loop antenna made of copper and isolated using a 1mm. teflon
like coating. Sea water is modelled as a dielectric with permittivity 
$\epsilon_r$=81 and conductivity $\sigma$=4.5~S/m. Seabed, fine sand, is 
modelled
as a dielectric with permittivity $\epsilon_r$=3.5 and conductivity
$\sigma$=1~S/m~\cite{cella:2009}. Height of sea water layer is set to \emph{h}=4~m.
\par
Simulations are carried out using a commercial MoM solver: FEKO. This tool
supports the features needed for this analysis: planar Green functions for 
multilayered media, dielectric coated wires and special basis functions for low
frequency analysis.
\par
Simulations are carried out at five different distances (\emph{d}=2, 3, 4, 5 and
6 meters) with frequency sweeps from 10~kHz to 1~MHz.

\section{Simulation results}
\subsection{Antennas inmersed in homogeneous medium}
\begin{figure}[!h]
\centering
\includegraphics[width=8.5cm]{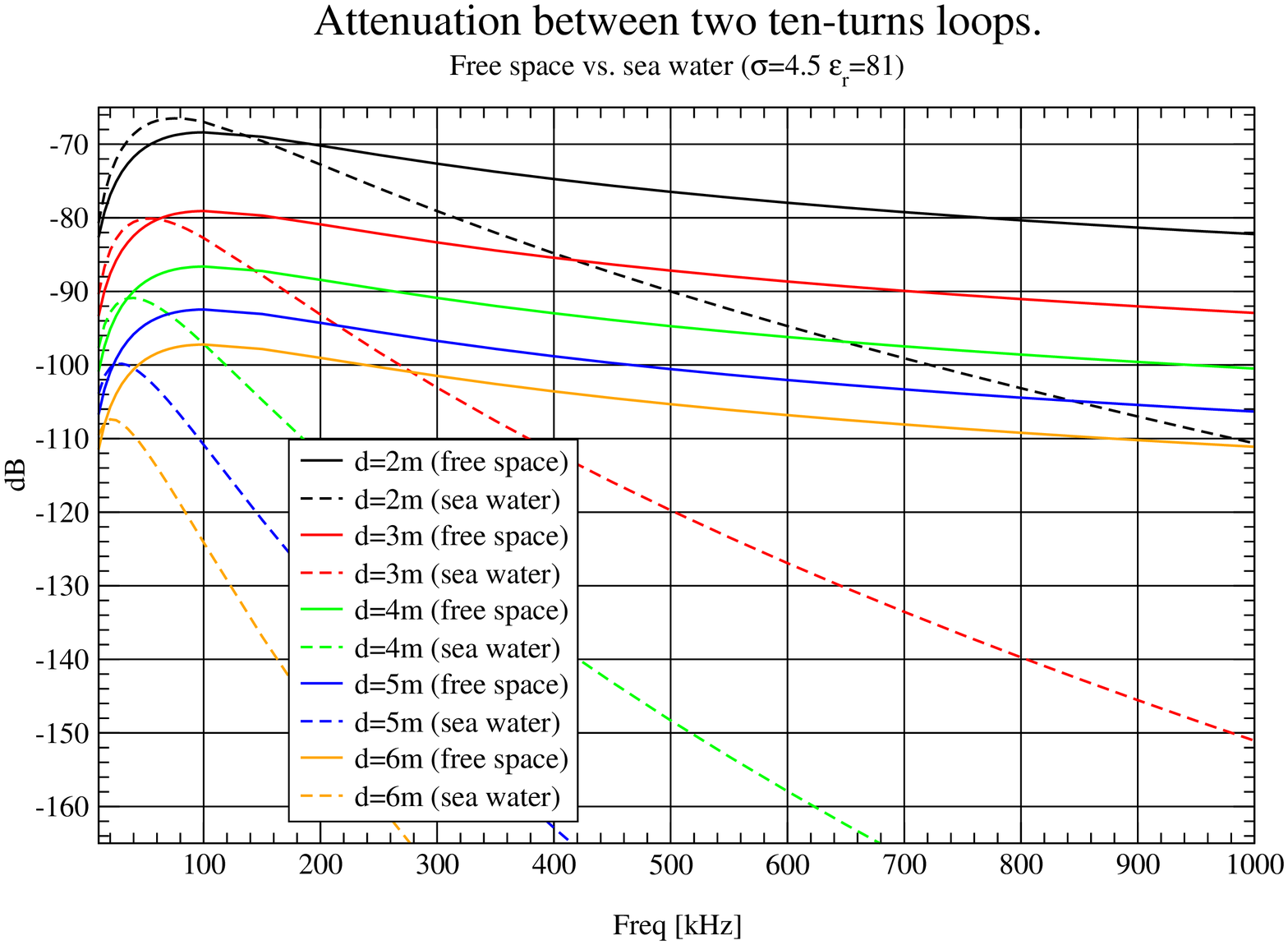}
\caption{Free space vs. sea water. Full sweep.}
\label{fig:air_sea}
\end{figure}

\begin{figure}[!h]
\includegraphics[width=8.5cm]{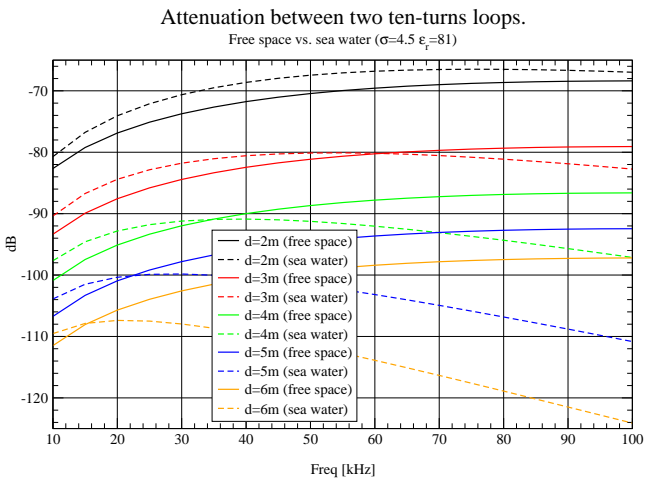}
\caption{Free space vs. sea water. Low frequency sweep.}
\label{fig:air_sea_low}
\end{figure}

In these simulations, antennas are radiating into two homogeneous mediums: free
space and sea water. Results of those simulations are shown 
in Fig~\ref{fig:air_sea} and in Fig~\ref{fig:air_sea_low}.
\par
As it is expected for antennas
inmersed in sea water, attenuation grows exponentially with frequency due to 
sea water conductivity (attenuation constant 
$\alpha = \sqrt{\frac{\sigma \omega \mu}{2}}$~Neper/m).
This is true for frequencies over
100~kHz but not so true for frequencies between 10~kHz and 100~kHz. For low
frequencies and distances or for very low frequencies, attenuation decreases 
with frequency in both cases:
free space and sea water, and it seems to be independent of the electrical
properties of the medium. In this case a magnetostatic approach can explain this
behaviour.
\par
For antennas in free space and frequencies over 100~kHz, 
atenuattion for each frequency increases 18~dB when doubling distance. It is a
typical near field dependence (eg $1/R^3$).
\par 
Electrically small loop antennas work as vertical magnetic dipoles.
The electromagnetic fields generated by this source~\cite{king:1992}
are (cylindrical coordinates):
\begin{IEEEeqnarray*}{l}
B_{\rho}=\frac{1}{4\pi\omega}\frac{\rho z}{r^2}
\left(\frac{jk^2}{r}-\frac{3k}{r^2}-\frac{3j}{r^3}\right)e^{jkr}\\
B_z=-\frac{1}{4\pi\omega}\left[\frac{jk^2}{r}-\frac{k}{r^2}-\frac{j}{r^3}
-\frac{z^2}{r^2}\left(\frac{jk^2}{r}-\frac{3k}{r^2}-\frac{3j}{r^3}\right)
\right]e^{jkr}
\\
E_{\phi}=-\frac{1}{4\pi}\frac{\rho}{r}
\left(\frac{jk}{r}-\frac{1}{r^2}\right)e^{jkr}
\end{IEEEeqnarray*}
\par
These equations clearly show the aforementioned behaviours: mainly magnetic
field for low frequencies and distances and $1/R^3$ near field dependence for 
low distances.
\subsection{Antennas inmersed in layered medium}
Now we are going to compare the results from simulation of antennas in sea water
with simulations of antennas placed on seabed. Both layers, sea water and
seabed, are semi-infinite so it is a two layer geometry.
\par
\begin{figure}[!h]
\centering
\includegraphics[width=8.5cm]{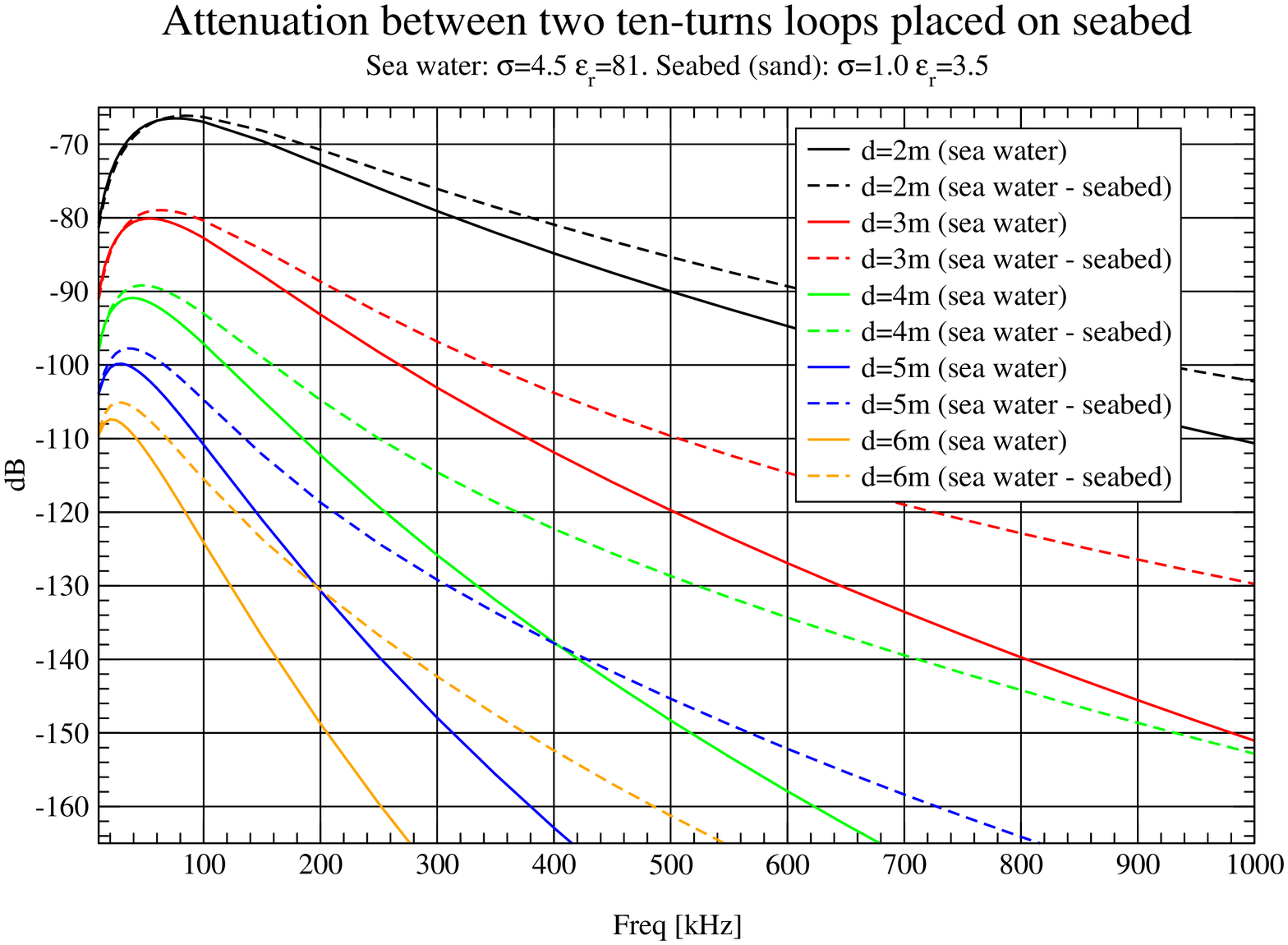}
\caption{Sea water vs. two layer. Full sweep.}
\label{fig:sea_2layer}
\end{figure}
\begin{figure}[!h]
\includegraphics[width=8.5cm]{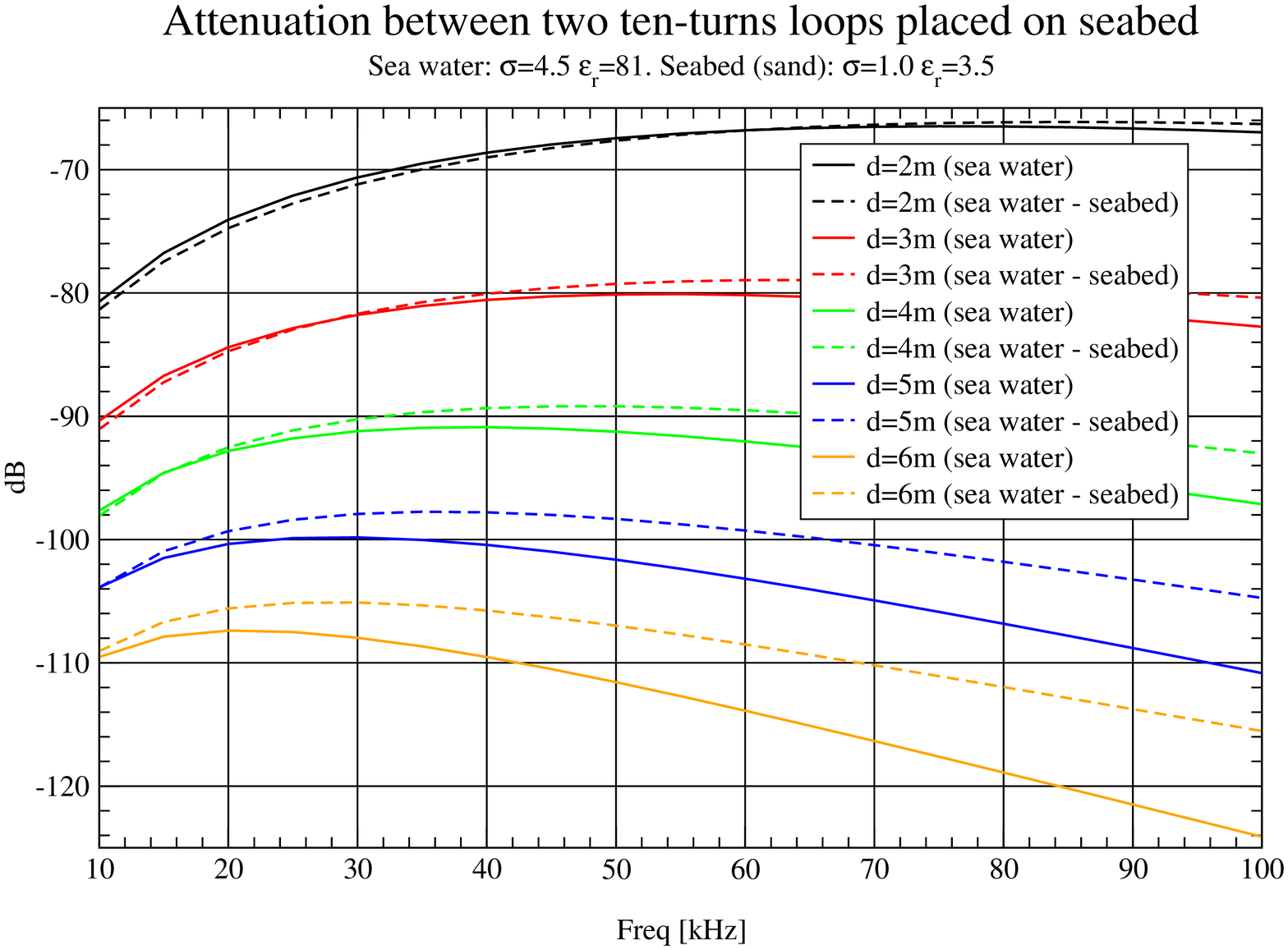}
\caption{Sea water vs. two layer. Low frequency sweep.}
\label{fig:sea_2layer_low}
\end{figure}

Results of those simulations are shown 
in Fig~\ref{fig:sea_2layer} and in Fig~\ref{fig:sea_2layer_low}.
\par
As it can be seen, the effect of seabed layer is to decrease the attenuation at
all frequencies. This means that a surface wave, \emph{lateral wave}, has been
launched and energy is mainly travelling on the sea-seabed interface.
Once again, for low frequencies and distances or for very low frequencies, 
attenuation seems to be independent from the medium.
\par
Full expressions for the fields generated by a vertical dipole placed on the
interface between two mediums can be found in~\cite{king:1992} and 
in~\cite{margetis:2001}. We are not going to reproduce them here because of
their length and complexity.  
\par
Simulations with three layers (seabed, sea water and air) has been carried out 
too. Results of those simulations, with water heigh \emph{h}=4m, are
indistinguishable from those of the two layers. So, at least for our testbed,
the seawater to air interface seems not to have any effect.

\section{Experimental results}
After reviewing a great number of studies about underwater propagation, we have
found little information about experimental results in this frequency band.
Therefore, a measurement system was designed and several experiments were
carried out along 2015 and 2016. After debugging a lot of problems we came to a
conclusion: the only way to measure without interferences in this band was to
\textbf{submerge all of the equipment in the sea} and communicate with it using
a fiber link. No copper cables from undersea to ground, even coaxial ones work
like antennas!
\par
\begin{figure}[!h]
\centering
\includegraphics[width=8.5cm]{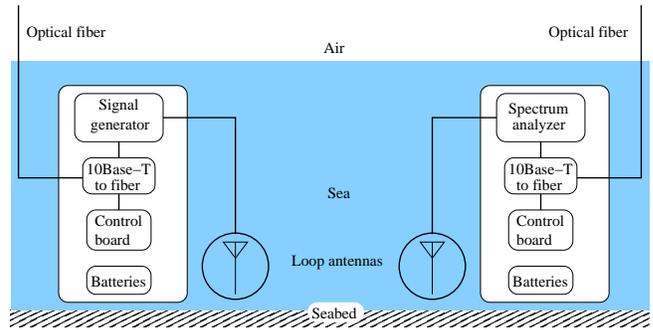}
\caption{Experimental testbed setup.}
\label{fig:testbed}
\end{figure}
\begin{figure}[!h]
\centering
\includegraphics[height=6.5cm]{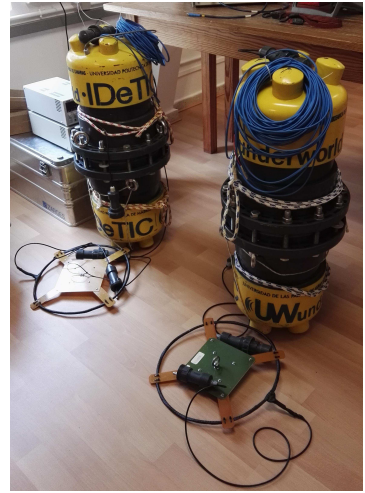}
\caption{Transmitter and receiver.}
\label{fig:minioms}
\end{figure}

The selected location is in Taliarte Harbour (Telde, Canary Islands, Spain).
This location was selected because PLOCAN's headquarters are placed there and we
can use a private pier. The testbed is shown in Fig~\ref{fig:testbed} and a
photograph of both systems is shown in Fig~\ref{fig:minioms}.
\par
A full description and details of the design of the experimental seabed can be
found in~\cite{pablo:2016}.
\par
\subsection{Transmitter and antenna}
The transmitter is built using a Keysight 33220A waveform generator, a
Beaglebone Black board, a 10Base-T to fiber transceiver and a battery pack with
an inversor. All the equipment is placed into a receptacle made of high 
pressure PVC
pipe. The loop antenna is made of enamelled coper covered with self-vulcanizing
tape and it ts conected to the transmitter using a short patch of coaxial line.
The PVC receptacle is pressurized and the control board sends information about
pressure and temperature using the fiber link. The generator is controlled from
an external computer using Keysight VEE software.
\subsection{Receiver and antenna}
The receiver is built using a handheld Keysight 9340B spectrum analyzer, a
Beaglebone Black board, a 10Base-T to fiber transceiver and a battery pack. 
All the equipment is placed into a receptacle made of high 
pressure PVC pipe. The loop antenna is the same used in the
transmitter and the PVC receptacle is pressurized too. The analyzer is controlled from
an external computer using Keysight VEE software.
\par
\subsection{Results}
Frequency sweeps were made between 10~kHz and 100~kHz (1~kHz IF bandwidth) and
between 100~kHz and 1~MHz (3~kHz IF bandwidth). Both antennas were placed on the
seabed and the distance between their centers was swept between 2 and 6 meters
using one meter steps. Signal generator power was set to 18 dBm for distances between 2
and 5 meters. For 6 meters, signal generator power was set to 23 dBm.
\par
After making the measurements, data from the spectrum analyzer needs to be
calibrated with a well known source. The spectrum analyzer has a poor response 
below 40~kHz (it's rated for use from 100~kHz) and
calibration curves have to be made in the lab to improve its response. These
curves help extracting the effects of analyzer in the measurements.
\par
\begin{figure}[!h]
\centering
\includegraphics[width=8.5cm]{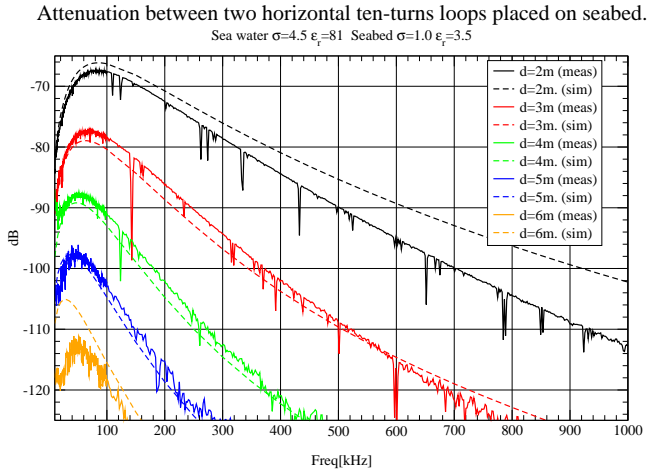}
\caption{Measurements vs. simulations. Full sweep.}
\label{fig:sim_meas}
\end{figure}

In Fig~\ref{fig:sim_meas} a full sweep between 10~kHz and 1~MHz is shown.
In this figure, simulations of a two layer model (water-seabed) are compared
with measurements. Heigth of water was four meters during the measurements.
\par
\begin{figure}[!h]
\includegraphics[width=8.5cm]{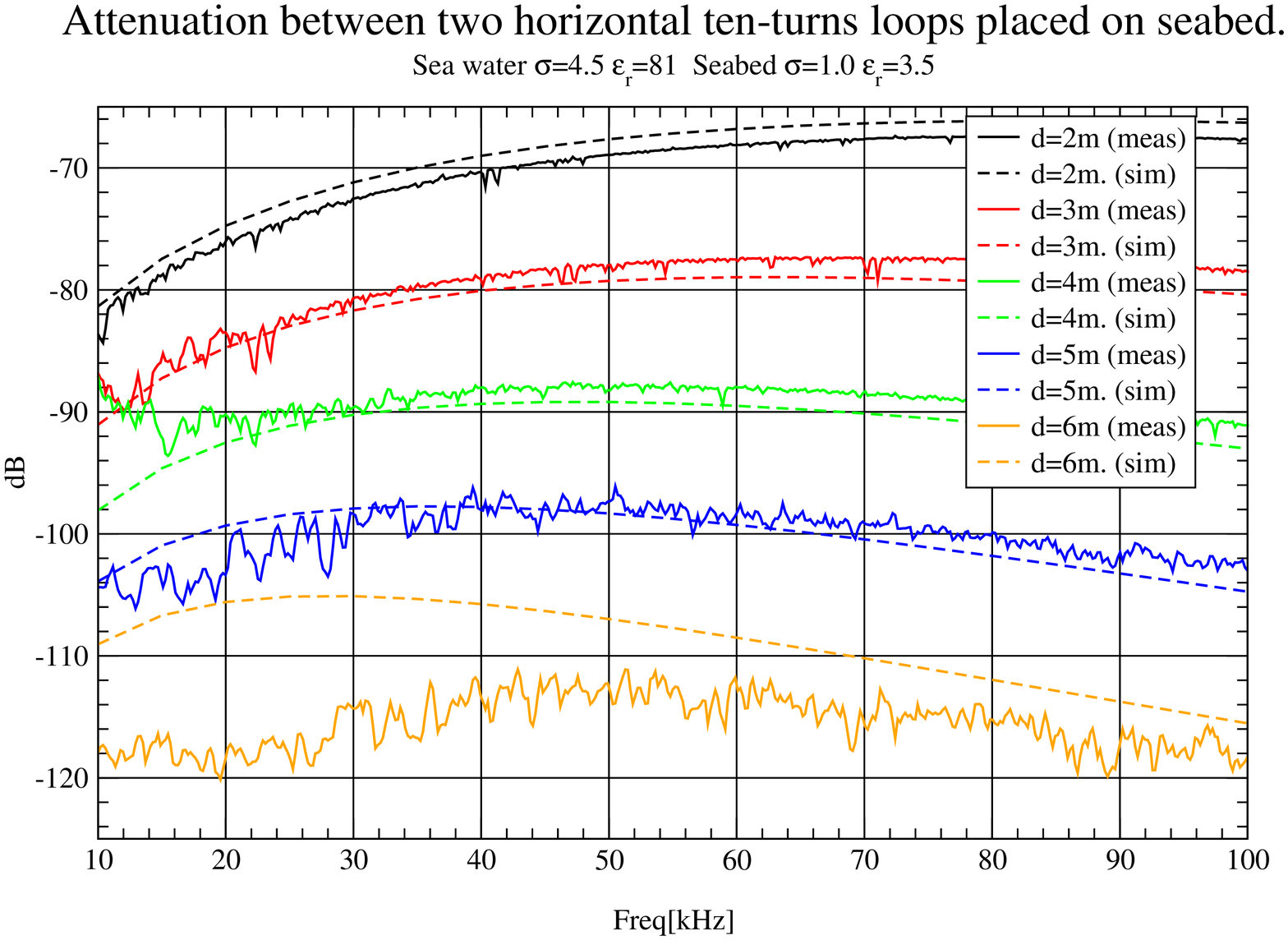}
\caption{Measurements vs. simulations. Low frequency sweep.}
\label{fig:sim_meas_low}
\end{figure}

In Fig~\ref{fig:sim_meas_low} a low frequency sweep between 10~kHz and 100~kHz is 
shown. Results for 4, 5 and 6 meters and for frequencies below 40~kHz are
largely influenced by the response of the spectrum analyzer at low power levels
at these frequencies. A new spectrum
HF analyzer from Aaronia GmbH (1~Hz to 30~Mhz) has been acquired and a new
measurement campaign is being planned as of writing this paper.

\section{Conclusions}
This paper investigates the propagation of EM waves generated by
loop antennas horizontally placed on seabed at
different frequencies and distances. Full wave simulations and measurements are
carried out for the same testbed geometry and two conclusions can be drawn.
\par
First, simulations predict that for frequencies over 100~kHz propagation takes
place mainly on the seabed-seawater interface. The simulated attenuation is 
greater in an homogeneous medium (seawater) than in a two layer medium
(seawater-seabed). These predictions agreed with the measurements.
\par
Second, simulations predict that for low frequencies the influence of the medium
decreases with the frequency showing a behaviour that can be explained using a
magnetostatic approach. This effect is stronger at short distances. These
predictions agree with the measurements too.
\par
The good agreement between simulations and measurements validates the simulation
tool. It will let us to make "numerical" experiments with antenna placement
(vertical, horizontal, etc\dots), with antenna geometry (radius of the loop,
number of turns, shape, etc\dots) and with frequency choice.   
\section*{Acknowledgment}
The authors would like to thank the work carried out by Juan Domingo Santana
Urbin (ULPGC) when making the pressurized PVC receptacles, loop antennas and 
all of the
lab stuff needed  to make the measurements. Also, the authors would like to
thank the work carried out by Gabriel Juanes and Raul Santana (PLOCAN) 
when setting up measurement
testbed in the pier and into the sea.
\par
This work has been supported by Ministerio de Economia y Competitividad, Spain,
under public contract TEC2013-46011-C3-R.
\ifCLASSOPTIONcaptionsoff
  \newpage
\fi

\bibliographystyle{IEEEtran}
\bibliography{IEEEabrv,ucomms}

\end{document}